%% file: bad470.tex
\documentclass[11pt]{article}
\usepackage{graphicx,amsmath}

\newcommand{\bzjpsipiz}{\ensuremath{\Bz\to \jpsi\piz}}

\newcommand{\bzjpsikspizpiz}{\ensuremath{\Bz\to\jpsi\KS(\piz\piz)}}
\newcommand{\De} {$\mathrm{\Delta}E$}
\newcommand{\DT} {\mbox{$\Delta t$}\xspace}
\newcommand{\sdt} {$\mathrm{\sigma_{\Delta t}}$}

\newcommand{\BABARPubYear}    {02}

\newcommand{\BABARConfNumber} {015}
\newcommand{\SLACPubNumber} {9298}

\input pubboard/babarsym

\long\def\inst#1{\par\nobreak\kern 4pt\nobreak
    {\it #1}\par\vskip 10pt plus 3pt minus 3pt}

\begin{document}
{\pagestyle{empty}


\begin{flushright}
\babar-CONF-\BABARPubYear/\BABARConfNumber \\
SLAC-PUB-\SLACPubNumber \\
July 2002 \\
\end{flushright}

\par\vskip 5cm

\begin{center}
\Large \bf A Study of Time-Dependent {\boldmath ${C\!P}$} Asymmetry in {\boldmath
  ${B^0\rightarrow J\mskip -3mu/\mskip -2mu\psi\mskip 2mu \pi^0}$} Decays
\end{center}
\bigskip

\begin{center}
\large The \babar\ Collaboration\\
\mbox{ }\\
July 24, 2002
\end{center}
\bigskip \bigskip

\begin{center}
\large \bf Abstract
\end{center}
  We present our first study of the time-dependent ${C\!P}$-violating
  asymmetry in ${B^0\rightarrow J\mskip -3mu/\mskip -2mu\psi\mskip
  2mu \pi^0}$ decays using ${e^+e^-}$ annihilation data collected with the
  ${\mbox{\sl B\hspace{-0.4em} {\small\sl A}\hspace{-0.37em} 
  \sl B\hspace{-0.4em} {\small\sl A\hspace{-0.02em}R}}}$ detector at the
  ${\Upsilon{\rm( 4S)}}$ resonance during the years 1999--2002 at the
  PEP-II asymmetric-energy $B$ Factory at SLAC.
  With about $88$ million \BB\ pairs, our preliminary results for
  the coefficients of the cosine and sine
  terms of the ${C\!P}$ asymmetry are $C_{\jpsi\piz} = 0.38 \pm 0.41\
  \stat \pm 0.09\ \syst$ and $S_{\jpsi\piz} = 0.05 \pm 0.49\ \stat
  \pm 0.16\ \syst$.

\vfill
\begin{center}
Contributed to the 31$^{st}$ International Conference on High Energy Physics,\\ 
7/24---7/31/2002, Amsterdam, The Netherlands
\end{center}

\vspace{1.0cm}
\begin{center}
{\em Stanford Linear Accelerator Center, Stanford University, 
Stanford, CA 94309} \\ \vspace{0.1cm}\hrule\vspace{0.1cm}
Work supported in part by Department of Energy contract DE-AC03-76SF00515.
\end{center}

\newpage
} 

\input pubboard/authors_ICHEP2002.tex

\section{Introduction}
\label{sec:Introduction}
The Standard Model of electroweak interactions describes
${C\!P}$-violation in \B\ meson decays by a complex phase in the
three-generation Cabibbo-Kobayashi-Maskawa (CKM)~\cite{ref:CKM} quark-mixing
matrix.  The ${b \rightarrow c\mskip 2mu \overline c \mskip
2mu s}$ modes that decay through charmonium, such as
${B^0 \rightarrow J\mskip -3mu/\mskip -2mu\psi\mskip 2mu 
K^0_{\scriptscriptstyle S}}$, yield precise measurements of
the quantity $\stwob$, where $\beta \equiv \arg \left[\,
  -V_{\rm cd}^{}V_{\rm cb}^* / V_{\rm td}^{}V_{\rm tb}^*\, \right]$
(see for example Refs.~\cite{ref:BABARs2b, ref:babar_s2b_new_prl, ref:BELLEs2b}).
The decay \bzjpsipiz\ is a Cabibbo-suppressed
${b \rightarrow c\mskip 2mu \overline c \mskip 2mu d}$
decay, whose tree contribution has the same weak phase as the
${b \rightarrow c\mskip 2mu \overline c \mskip 2mu s}$ modes
(e.g.\ ${B^0 \rightarrow J\mskip -3mu/\mskip -2mu\psi\mskip 2mu 
K^0_{\scriptscriptstyle S}}$).  A portion of the penguin 
contribution has a different weak phase, which may give 
a time-dependent ${C\!P}$ asymmetry that differs from the one
observed in ${b \rightarrow c\mskip 2mu \overline c \mskip 2mu s}$
decays.

In this measurement, about $88$ million \FourS\to\BB\ decays are used to detect
the decay chain \bzjpsipiz, with \jpsi\to\epem\ or \jpsi\to\mumu.
The \babar\ measurement of the \bzjpsipiz\ branching fraction, $(2.0
\pm 0.6\ \stat \pm 0.2\ \syst) \times 10^{-5}$, is described elsewhere~\cite{ref:excl_B_PRD}.
Properties of the recoiling
\B\ meson are used to infer the flavor (\Bz\ or \Bzb) of the
\B\ meson that is reconstructed from \jpsi\ and \piz\ candidates.  The
decay time distribution of $B$ decays to a ${C\!P}$ eigenstate with a \Bz
or \Bzb flavor tag can be expressed in terms of a complex parameter $\lambda$
that depends on both the \Bz-\Bzb oscillation amplitude and the amplitudes
describing \Bzb and \Bz decays to this final
state~\cite{ref:lambda}. The decay rate  ${\rm f}_+({\rm f}_-)$ when the 
tagging meson is a $\Bz (\Bzb)$ is given by 

\begin{eqnarray}
{\rm f}_\pm(\, \deltat) = {\frac{{e}^{{- \left| \deltat \right|}/\tau_{\Bz} }}{4\tau_{\Bz}
}}  \times  \left[ \ 1 \hbox to 0cm{}
\pm \frac{{2\mathop{\cal I\mkern -2.0mu\mit m}}
\lambda}{1+|\lambda|^2}  \sin{( \deltamd  \deltat )} 
\mp { \frac{1  - |\lambda|^2 } {1+|\lambda|^2} }  
  \cos{( \deltamd  \deltat) }   \right],
\label{eq:timedist}
\end{eqnarray}

\vskip12pt\noindent
where $\Delta t = t_{\rm rec} - t_{\rm tag}$ is the difference between
the proper decay time of the reconstructed $B$ meson ($B_{\rm rec}$) and 
the proper decay time of the tagging $B$ meson ($B_{\rm tag}$),
$\tau_{\Bz}$ is the \Bz lifetime, and \deltamd is the \Bz-\Bzb oscillation
frequency.  The sine
term in Eq.~\ref{eq:timedist} is due to the interference between direct
decay and decay after flavor change, and the cosine term is due to the
interference between two or more decay amplitudes with different weak
and strong phases.  Two amplitudes can contribute in the decay
\bzjpsipiz.  A portion of the penguin amplitude has the
same weak phase as the tree amplitude, while the remainder of the
penguin amplitude has a different weak phase.
In ${b \rightarrow c\mskip 2mu \overline c \mskip 2mu d}$ decays, the
tree contribution is Cabibbo-suppressed
and the penguin and tree diagrams may enter at the same order,
proportional to $\lambda^3$ (where in this case $\lambda$ is the
Wolfenstein parameter of the CKM matrix, rather than the
complex parameter that appears in Eq.~\ref{eq:timedist}).
Evidence for ${C\!P}$ violation can be observed as a difference between the
\deltat distributions of \Bz- and \Bzb-tagged events or as
an asymmetry with respect to $\deltat = 0$ for either flavor tag.
We measure the two asymmetry coefficients, defined as

\begin{eqnarray}
S_{f} \equiv \frac{{2\mathop{\cal I\mkern -2.0mu\mit m}}
  \lambda}{1+|\lambda|^2}
\mskip 50mu {\rm and} \mskip 50mu
C_{f} \equiv \frac{1  - |\lambda|^2 }{1+|\lambda|^2},
\label{eq:coef_def}
\end{eqnarray}

\noindent
where $f$ is the final state.  With these definitions, the absence of
penguin contributions would give $S_{\jpsi\piz} = -\stwob$ and $C_{\jpsi\piz} = 0$.
A statistically significant deviation from these values may indicate
penguin contributions not only in \bzjpsipiz, but also in ${B^0
  \rightarrow J\mskip -3mu/\mskip -2mu\psi\mskip 2mu
  K^0_{\scriptscriptstyle S}}$ (at a reduced level governed by
Cabibbo suppression).

\section{The \babar\ detector and dataset}
\label{sec:babar}
The data used in this measurement were collected with the \babar\ detector
at the \pep2\ storage ring from 1999 to 2002.  Approximately
$81\invfb$ of ${e^+e^-}$ annihilation data taken at the
\FourS\ resonance are used, corresponding to
a sample of about $88$ million \BB\ pairs.  An additional $5\invfb$ of
data collected approximately $40 \mev$ below the \FourS\ resonance are
used to characterize one of the background sources.

The \babar\ detector is described in detail elsewhere~\cite{ref:babar}.  
Surrounding the beam pipe is a silicon vertex tracker (SVT), which
provides precise measurements of the trajectories of charged particles
as they leave the \epem interaction point.  
A 40-layer drift chamber (DCH) surrounds the SVT, and both
allow measurements of track momenta in a 1.5-T magnetic field as well as
energy-loss measurements, which contribute to charged
particle identification. Surrounding the DCH is a detector of internally
reflected Cherenkov radiation (DIRC), which provides charged hadron
identification. Outside of the DIRC is a CsI(Tl) electromagnetic
calorimeter (EMC) that is used to detect photons, provide electron
identification, and reconstruct neutral hadrons. The EMC is surrounded by the
superconducting coil, which creates the magnetic field for momentum
and charge measurements.  Outside of the coil, the 
flux return yoke is instrumented with resistive plate chambers interspersed with 
iron (IFR) for the identification of muons and long-lived neutral hadrons.

\section{Candidate selection}
\label{sec:selection}
\bzjpsipiz\ candidates are selected by
identifying \jpsi\to\epem\ or
\jpsi\to\mumu\ decays and \piz\to\gaga\ decays (details are given in
Ref.~\cite{ref:excl_B_PRD}).  For the
\jpsi\to\epem\ channel, photons consistent with bremsstrahlung are
added and each lepton candidate must be consistent with the electron
hypothesis.
For the \jpsi\to\mumu\ channel, each lepton candidate must
be consistent with the muon hypothesis.  The invariant mass of
the lepton pair is required to be between $2.95$ and $3.14 \gevcc$,
and $3.06$ and $3.14 \gevcc$, for the electron and muon channels,
respectively.
The photon candidates used to reconstruct the \piz\ candidate are
identified as clusters in the EMC within the polar angle
range $0.410 < \theta_{\rm lab} < 2.409 \rad$ that are spatially separated from
every charged track, and have a minimum energy of $30 \mev$.
The lateral energy distribution in the cluster is required to be
consistent with a photon.  The invariant mass of the photon pair is
required to be $100 < m_{\gaga} < 160 \mevcc$.
Finally, the \jpsi\ and \piz\ candidates defined above are combined
using a mass-constrained kinematic vertexing algorithm.

Two kinematic consistency requirements are applied to each \B\ candidate.
The difference, \De, between the \B\ candidate energy and the
beam energy in the center-of-mass frame must be $-0.4 <
\mathrm{\Delta}E < 0.4 \gev$.  The beam-energy-substituted mass,
$\mes=\sqrt{{(E^{\rm  *}_{\rm beam})^2}-(p_B^{\rm *})^2}$, must be
$5.2 < \mes\ < 5.3 \gevcc$, where $E^{\rm  *}_{\rm beam}$ and
$p_B^{\rm *}$ are the beam energy and \B\ candidate momentum in the
center-of-mass frame.

Several kinematic and topological variables are linearly combined
using a Fisher discriminant, $\cal F$,
to provide additional separation between signal and
$\epem\to\uubar,\ddbar,\ssbar,\ccbar$ (continuum) background events.
The inputs to the Fisher discriminant are:  the zeroth and second order
Legendre polynomial momentum moments ($L_0 = \sum_i |{\bf p}_i|$ and
$L_2 = \sum_i |{\bf p}_i| \hspace{0.5mm} \frac{3 \cos^2\theta_i - 1}{2}$,
where ${\bf p}_i$ are the momenta for the charged and neutral objects
in the event that are not associated with the signal candidate, and
$\theta_i$ are the angles between ${\bf p}_i$ and the thrust axis of
the signal candidate);  the ratio of the
second-order to zeroth-order Fox-Wolfram moment~\cite{ref:fox_wolf}, computed using all
charged and neutral objects not associated with the signal candidate;
$|\cos{\theta_T}|$, where $\theta_T$ is the angle between the thrust
axis of the \B\ candidate and the thrust axis of the
remaining charged tracks and neutral objects in the event;
$|\cos{\theta_\ell}|$, where $\theta_\ell$
is the lepton helicity angle, defined as the angle between the
negative lepton and \B\ candidate directions in the
\jpsi\ rest frame.  We require ${\cal F} > -0.8$, which is $99\%$
efficient for signal and rejects $71\%$ of the continuum background.
The efficiencies for satisfying this requirement are summarized in
Table~\ref{table:efficiencies}.

\begin{table}
\small
\caption{The efficiencies for the requirement on the Fisher discriminant
  and tagging, given independently, with statistical uncertainties.}
\begin{center}
\begin{tabular}{|l|c|c|} \hline
Source type        & Efficiency (\%) of ${\cal F} > -0.8$ & Tagging efficiency (\%) \\ \hline \hline
\bzjpsipiz\                                & $99.2 \pm 0.1$ & $65.6 \pm 0.6$ \\ \hline
\bzjpsikspizpiz\ background                & $98.9 \pm 0.1$ & $65.6 \pm 0.6$ \\ \hline
\B\to\jpsi\ X (inclusive \jpsi) background & $94.9 \pm 0.7$ & $70.4 \pm 1.4$ \\ \hline
\B\to\ X (\BB\ generic) background         & $98.5 \pm 0.4$ & $61.1 \pm 1.6$ \\ \hline
\epem\to\qqbar (continuum) background      & $28.6 \pm 0.7$ & $52.3 \pm 0.8$ \\ \hline
\end{tabular}
\end{center}
\label{table:efficiencies}
\end{table}

\section{Backgrounds}
\label{sec:backgrounds}
The backgrounds come from decays which contain a \jpsi\ particle or
from purely random combinations.  We split the backgrounds into four categories.

One of the \B\to\jpsi\ X decays is \bzjpsikspizpiz.
In this case, one of the \piz's is emitted nearly at rest in the center-of-mass
frame, and is thus missed in the reconstruction of the \B\ candidate.
The second is the more general class of \B\to\jpsi\ X (inclusive
\jpsi) decays, which contribute through random combinations of \jpsi\ and
\piz\ candidates.  This also includes cascade
decays through other charmonium states, but excludes the specific
\bzjpsikspizpiz\ mode discussed above.
Third is a purely combinatoric background contribution coming from the
general decay \B\to\ X (\BB\ generic).  Excluded from this definition
are those decays already considered above.
The fourth type of background is a combinatoric background due to
\u, \d, \s, and \c quark production
following the \epem\ annihilation, \epem\to\qqbar (continuum).
We study this background using an inverted lepton particle
identification selection on the below-resonance data sample.  In this
case, the \jpsi\ candidate is reconstructed from two particle
candidates that are not consistent with a lepton hypothesis.  Monte
Carlo simulation is used to check that this procedure correctly models
the background.

\section{Flavor tagging and measurement of {\boldmath \DT}}
\label{sec:tagging_and_vertexing}
The methods for \B\ flavor tagging, vertex reconstruction, and the
determination of \DT, are described in
Refs.~\cite{ref:babar_s2b_new_prl, ref:babar_s2b_PRD}.
For flavor tagging, we exploit information from the recoil \B\ decay in
the event. The charges of energetic 
electrons and muons from semileptonic \B\ decays, kaons, soft pions from
\Dstarp\to\Dz\pip\ decays, and  high momentum particles are correlated with
the flavor of the decaying \B\ meson. 
For \B\ decays, about $66\%$ of the events can be assigned to one of
four hierarchical, mutually exclusive tagging categories.  The
remaining untagged events are excluded from further
analysis.  The total tagging efficiency for each source type is shown
in Table~\ref{table:efficiencies}.

The time interval \DT\ between the two \B\ decays is calculated
from the measured separation $\Delta z$ between the decay vertex of the 
reconstructed  \B\ meson ($B_{\rm rec}$) and the vertex of the
flavor-tagging \B\ meson ($B_{\rm tag}$) along the beam axis ($z$ axis). 
The calculation of \DT\ includes an event-by-event
correction for the direction of the $B_{\rm rec}$ with respect to  
the $z$ direction in the $\FourS$ frame. We determine the $z$ position
of the $B_{\rm rec}$ vertex from the reconstructed vertex of the
\jpsi\ candidate. The
$B_{\rm tag}$ vertex is determined by fitting the tracks not
belonging to the $B_{\rm rec}$ candidate to a common vertex. An
additional constraint on the tagging vertex comes from a pseudo-track 
computed from the  $B_{\rm rec}$ vertex and three-momentum,
the beam-spot (with a vertical size of $10 \mum$), and the \FourS
momentum. For $99.5\%$ of the reconstructed events the r.m.s.\ $\Delta z$
resolution is $180\mum$. Convergence is required for both the
$B_{\rm rec}$ and $B_{\rm tag}$ vertex fits.  Finally,
\DT\ must be between $-20$ and $20 \ps$, and it is required to
have an uncertainty satisfying $\sigma_{\Delta t} < 2.4$ ps.

\section{Maximum likelihood fitting technique}
\label{sec:likelihood_fit}

We extract the ${C\!P}$ asymmetry by performing an unbinned extended
likelihood fit.  The likelihood is constructed from the probability density
functions for the discriminating variables \mes, \De, and \DT.
The quantity that is maximized is

\begin{align}
\label{eq:CP_likelihood}
{\cal L} = \frac{e^{-\sum_{j=1}^{5}n_j}}{N!}     \nonumber
        \prod_{i=1}^{N} \hspace{2mm}
    & \{ f^{Sig}_{\alpha} \hspace{2mm} n_{Sig}
    \hspace{2mm} {\cal P}^{Sig} _{\mbox{\mes}} \hspace{2mm} 
        {\cal P}^{Sig}_{\Delta E}
        \hspace{2mm} {\cal P}^{Sig}_{\Delta t} \\ \nonumber
    &  + f^{Ks}_{\alpha} \hspace{2mm} n_{Ks}
    \hspace{2mm} {\cal P}^{Ks}_{\mbox{\mes--}\Delta E} 
    \hspace{2mm} {\cal P}^{Ks}_{\Delta t} \\ \nonumber
    &  + f^{Inc}_{\alpha} \hspace{2mm} n_{Inc}
    \hspace{2mm} {\cal P}^{Inc}_{\mbox{\mes--}\Delta E} 
    \hspace{2mm} {\cal P}^{Inc}_{\Delta t} \\ \nonumber
    &  + f^{BB}_{\alpha} \hspace{2mm} n_{BB}
    \hspace{2mm} {\cal P}^{BB}_{\mbox{\mes}} 
    \hspace{2mm} {\cal P}^{BB}_{\Delta E} 
    \hspace{2mm} {\cal P}^{BB}_{\Delta t} \\
    &  + f^{Cont}_{\alpha} \hspace{2mm} n_{Cont}
    \hspace{2mm} {\cal P}^{Cont}_{\mbox{\mes}} 
    \hspace{2mm} {\cal P}^{Cont}_{\Delta E}
   \hspace{2mm} {\cal P}^{Cont}_{\Delta t} \}, \\ \nonumber
\end{align}

\noindent
where $n_j$ is the number of events for each of the $5$ hypotheses ($1$
signal and $4$ backgrounds) and  $N$ is the number of input events.
The ${\cal P}$ are the probability density functions (PDFs) for each
discriminating variable and signal or background type.
The parameters $f^j_{\alpha}$ are the tagging
efficiencies for each of the 4 tagging categories $\alpha$ and each of
the signal or background types $j$.
For the \bzjpsipiz\ signal and \bzjpsikspizpiz\ background, the values
of $f^j_{\alpha}$ are measured with a sample
($B_{\rm flav}$) of neutral \B\ decays to flavor eigenstates
consisting of the channels $D^{(*)-}h^+ (h^+=\pi^+,\rho^+$, and $a_1^+)$ and
$\jpsi\Kstarz (\Kstarz\to\Kp\pim)$~\cite{ref:babar_s2b_new_prl}.  For
the inclusive \jpsi\ background and
\BB\ generic background, they are measured with Monte Carlo
simulation~\cite{ref:geant4}, and for
the continuum background, they are measured with the inverted lepton particle
identification data sample.  We discuss the discriminating variables
(\mes, \De, and \DT) in the following sections.

\subsection{Probability density functions for {\boldmath \mes\ and \De}}
\label{subsec:mes}

The signal \mes\ distribution is modeled as the sum of two components.
The first is a modified Gaussian function which, for values less than
the mean, has a width parameter that scales
linearly with the distance from the mean.
The second component, accounting for less than $6\%$ of the
distribution, is an ARGUS function~\cite{argus}, which is a
phase-space distribution of the form
${\mbox{\mes}} \sqrt{(1 - \frac{{\mbox{\mes}}^2}{E_{beam}^2})}
\hspace{1mm} {\rm exp}(\xi(1 -
\frac{{\mbox{\mes}}^2}{E_{beam}^2}))$, with a kinematic cut-off at $E_{beam}
= 5.289 \gev$, and one parameter to fit in the 
exponential, $\xi$.
The signal \De\ distribution is modeled by the sum of a Crystal Ball
function~\cite{CB} and a second order polynomial.  The Crystal Ball
function is defined as

\begin{align}
\label{eq:crystal}
C(\mathrm{\Delta}E) & = \left\{ \begin{array}{ll}
    e^{-\frac{(\mathrm{\Delta}E -m)^2}{2\sigma^2}} & \mbox{if $\mathrm{\Delta}E > m - \alpha \sigma$} \\
    \frac{ (\frac{n}{\alpha})^n \hspace{1mm} e^{-\frac{\alpha^2}{2}} }
        { (\frac{m-\mathrm{\Delta}E}{\sigma} + \frac{n}{\alpha} - \alpha)^n } &
                           \mbox{if $\mathrm{\Delta}E \le m - \alpha \sigma$, } \\
  \end{array}
\right. \nonumber \\
\end{align}

\noindent
where $m$ is the position of the maximum.  The parameter
$\alpha$ determines the cross-over point from a Gaussian behavior to
a power-law, and is expressed in units of the peak width $\sigma$.
The parameter $n$ is a real number which enters
into the power-law portion of the function.
The parameters of these PDFs are determined by fitting to a signal Monte
Carlo sample.  The maximum of the \De\ distribution is a free parameter
of the full ${C\!P}$ likelihood fit to allow for EMC energy scale
uncertainties.

The kinematic variables \mes\ and \De\ are correlated for the
\bzjpsikspizpiz\ and the inclusive \jpsi\ backgrounds.  To account for
this, 2-dimensional PDFs are employed.  Variably binned interpolated
2-dimensional histograms of these variables are constructed from the
relevant Monte Carlo samples.

The \mes\ PDFs for the \BB\ generic and
continuum backgrounds are modeled by the ARGUS function
and the \De\ PDFs for these two backgrounds are modeled by a second order
polynomial.  The parameters for these PDFs are obtained from
the \BB\ generic Monte Carlo sample and the inverted lepton particle
identification data sample.

\subsection{Probability density functions for {\boldmath \DT}}
\label{subsec:dt}

The PDFs used to describe the \DT\ distributions of the signal
and background sources are each a convolution of a resolution
function ${\cal R}$ and decay time distribution ${\cal D}$:
${\cal P}(\Delta t) = {\cal R}(\Delta t) \otimes {\cal D}(\Delta t)$.

For the signal, the resolution function~\cite{ref:babar_s2b_PRD}
consists of the sum of three Gaussians, which will be referred to 
as the core, tail, and outlier.
The means of the Gaussians are biased away from
zero due to the charm content of the side of the event used for
tagging.  For the core and tail
Gaussians this bias is multiplied by the \DT\ per-event error \sdt.
The widths of the core and tail Gaussians are the products of the
per-event errors and scale factors.  The tail Gaussian has a fixed scale
factor of $3$ and the outlier Gaussian has a fixed width of
$8 \ps$ and zero mean.  The five remaining parameters are measured with
the large $B_{\rm flav}$ data sample.  The bias of the
core Gaussian has different values for each of the four
tagging categories.

The decay time distribution is given by Eq.~\ref{eq:timedist} modified
for the effects of \B\ flavor tagging:

\begin{align}
\label{eq:dt_sig_D}
{\cal D}_{\alpha,f}^{\pm}(\deltat) = {\frac{{e}^{{- \left| \deltat
        \right|}/\tau_{\Bz} }}{4\tau_{\Bz}}} 
\{ & (1 \mp \Delta w_{\alpha})
\pm S_{f} \hspace{1mm} (1 - 2w_{\alpha}) \sin(\deltamd  \deltat)
\nonumber \\
& \mp C_{f} \hspace{1mm} (1 - 2w_{\alpha}) \cos(\deltamd \deltat)
 \}, \\ \nonumber
\end{align}

\noindent
where ${\cal D}_{\alpha,f}^{+}$(${\cal D}_{\alpha,f}^{-}$) is for a $\Bz$($\Bzb$) tagging meson.
The variable $w_{\alpha}$ is the average probability of
incorrectly tagging a \Bz\ as a \Bzb\ ($w^{\Bz}_{\alpha}$) or
a \Bzb\ as a \Bz ($w^{\Bzb}_{\alpha}$), and 
$\Delta w_{\alpha} = w^{\Bz}_{\alpha} - w^{\Bzb}_{\alpha}$.
Both $w_{\alpha}$ and $\Delta w_{\alpha}$ are determined using the
$B_{\rm flav}$ data sample~\cite{ref:babar_s2b_new_prl}.
The values of $\deltamd$ and $\tau_{\Bz}$
are the 2002 PDG averages~\cite{2002PDG}.

The PDF used to model the \DT\ distribution for the
\bzjpsikspizpiz\ background takes the same form as that for signal,
but with $S_{\jpsi\KS} = 0.75$~\cite{s2bCONF},
and $C_{\jpsi\KS} = 0$.

The parameterizations of the \DT\ PDFs for the inclusive
\jpsi\ background and the \BB\ generic background consist of
lifetime and prompt components.  
The resolution function for each component is the sum of core and
outlier Gaussians, where the width of each core Gaussian is the product
of $\sigma_{\Delta t}$ and a scale factor.  Once again, the width and
mean of the outlier Gaussians are fixed to $8 \ps$ and zero respectively.
For each of these background sources, the fraction
which is in the lifetime component, the decay lifetime parameter, and
the resolution parameters are the values determined from the Monte
Carlo simulation.

The \DT\ PDF for the continuum background consists of a double
Gaussian which has the same form as the prompt component of the
inclusive \jpsi\ and \BB\ generic \DT\ PDFs, where in this case the
parameter values are obtained by fitting the inverted lepton particle
identification data sample.

\subsection{Results of the {\boldmath ${C\!P}$} asymmetry fit}
\label{subsec:fit}

\begin{figure}
\begin{center}
\includegraphics[width=10.0cm,angle=0]{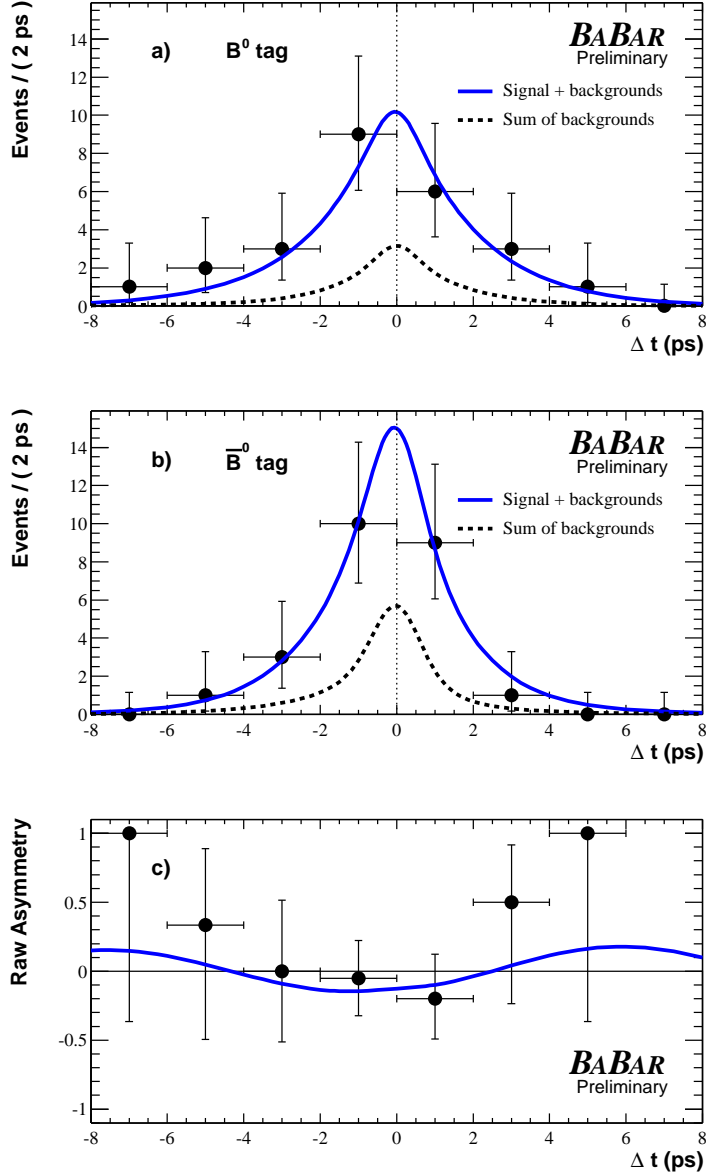}
\caption{Number of candidates in the signal region a) with a \Bz\ tag
  $N_{\Bz}$ and b) with a \Bzb\ tag $N_{\Bzb}$, and c) the raw asymmetry
  $(N_{\Bz} - N_{\Bzb})/(N_{\Bz} + N_{\Bzb})$, as functions of \DT.
  Candidates in these plots are required to satisfy
  $-0.11 < \mathrm{\Delta}E < 0.11 \gev$ and $\mes > 5.27 \gevcc$.
  The curves in a) and b) are projections that use the values of the
  other variables in the likelihood to determine the contribution to
  the signal or one of the backgrounds.
}
\label{fig:fit_asym}
\end{center}
\end{figure}

The results of the ${C\!P}$ asymmetry fit to 438 events
found in $81\invfb$ of data are shown in
Table~\ref{table:fit_results}.  The projections in \mes, \De, and \DT,
are shown in Figure~\ref{fig:fit_results}. Figure~\ref{fig:fit_asym}
shows the \DT\ distributions and asymmetries in yields between
\Bz\ and \Bzb\ flavor tags as functions of \DT, overlaid with the
projection of the likelihood fit results.

\begin{table}
\caption{Results of the ${C\!P}$ likelihood fit. Errors are
  statistical only.  The global correlation
  is $0.14$ for $C_{\jpsi\piz}$
  and $0.15$ for $S_{\jpsi\piz}$.  The projections of the PDFs
  are shown in Figure~\ref{fig:fit_results} and the asymmetry in
  Figure~\ref{fig:fit_asym}.}
\begin{center}
\begin{tabular}{|l|c|} \hline
                              & Fit results         \\ \hline \hline
$C_{\jpsi\piz}$                  & $0.38 \pm 0.41$  \\ \hline
$S_{\jpsi\piz}$                  & $0.05 \pm 0.49$  \\ \hline
Signal \De\ Maximum (\mev)          & $-13.2 \pm 7.2$\\ \hline
\bzjpsipiz\ signal          (Events)  & $40 \pm 7$      \\ \hline
\bzjpsikspizpiz\ background (Events)  & $140 \pm 19$    \\ \hline
Inclusive \jpsi\ background (Events)  & $109 \pm 35$    \\ \hline
\BB\ generic background     (Events)  & $52 \pm 25$     \\ \hline
Continuum background        (Events)  & $97 \pm 22$     \\ \hline
\end{tabular}
\end{center}
\label{table:fit_results}
\end{table}

\begin{figure}
\begin{center}
\includegraphics[width=14.55cm,angle=0]{cpfit_81.epsi}
\caption{Projections in a) \mes, c) \De, and e) \DT\ for the results
  of the ${C\!P}$ fit to $81\invfb$ of data.  The legend in a) applies
  to the plots on the left hand side.  The projection in b) \mes\ is
  shown with the requirement $-0.11 < \mathrm{\Delta}E < 0.11 \gev$.
  The projection in d) \De\ is shown with the requirement
  $\mes > 5.27 \gevcc$.  The projection in f) \DT\ is shown with the
  requirements $-0.11 < \mathrm{\Delta}E < 0.11 \gev$ and
  $\mes > 5.27 \gevcc$.  These plots do not represent the full
  information used in the maximum likelihood fit, but only a partial
  view of the data.}
\label{fig:fit_results}
\end{center}
\end{figure}

\section{Systematic uncertainties}
\label{sec:Systematics}

The contributions to the systematic errors in $C_{\jpsi\piz}$ and 
$S_{\jpsi\piz}$ are summarized in
Table~\ref{table:syst_error}.  The first class of uncertainties are
those obtained by variation of the parameters used in the \mes, \De,
and \DT\ PDFs, where the dominant sources are the uncertainties in the
signal \De\ PDF parameters.
Another contribution is due to the energy scale uncertainties in the
modeling of the \bzjpsikspizpiz\ background.
An additional systematic uncertainty comes from altering the
configuration of the 2-dimensional PDFs for the \bzjpsikspizpiz\ and
inclusive \jpsi\ backgrounds.
A systematic error to account for a correlation between
the tails of the signal \mes\ and \De\ distributions is obtained by
using a 2-dimensional PDF.

\begin{table}
\caption{Summary of the systematic errors.}
\begin{center}
\begin{tabular}{|l|c|c|} \hline
  Source                      & Error on $C_{\jpsi\piz}$ & Error on $S_{\jpsi\piz}$ \\ \hline \hline
\multicolumn{3}{|l|}{Parameter variations} \\ \hline
\,\,\,\, \mes\ and \De\ parameters            & $0.048$ & $0.130$ \\ \hline
\,\,\,\, Tagging fractions                    & $0.002$ & $0.007$ \\ \hline
\,\,\,\, \DT\ parameters                      & $0.027$ & $0.022$ \\ \hline
\multicolumn{3}{|l|}{Additional systematics} \\ \hline
\,\,\,\, EMC energy scale \bzjpsikspizpiz\    & $0.009$ & $0.002$ \\ \hline
\,\,\,\, Changing the 2-D histogram PDFs      & $0.009$ & $0.029$ \\ \hline
\,\,\,\, Using 2-D PDF for signal             & $0.073$ & $0.079$ \\ \hline
\,\,\,\, Beam spot, boost/vtx., SVT misalign. & $0.012$ & $0.012$ \\ \hline \hline
{\bf Total systematic uncertainty}   & $0.093$ & $0.157$ \\ \hline
\end{tabular}
\end{center}
\label{table:syst_error}
\end{table}

\section{Summary}
\label{sec:Summary}
An unbinned extended maximum likelihood fit has been performed on
$81\invfb$ of data collected at \babar, yielding preliminary values
for the coefficients of the cosine and sine terms of the time-dependent
${C\!P}$ asymmetry in \bzjpsipiz\ decays:

\begin{align}
C_{\jpsi\piz} & = 0.38 \pm 0.41\ \stat \pm 0.09\ \syst,\nonumber \\
S_{\jpsi\piz} & = 0.05 \pm 0.49\ \stat \pm 0.16\ \syst.\nonumber
\end{align}
\noindent

\section{Acknowledgments}
\label{sec:Acknowledgments}

\input pubboard/acknowledgements

\end{document}

%% file: pubboard/authors_ICHEP2002.tex
\begin{center}
\small

The \babar\ Collaboration,
\bigskip

B.~Aubert,
D.~Boutigny,
J.-M.~Gaillard,
A.~Hicheur,
Y.~Karyotakis,
J.~P.~Lees,
P.~Robbe,
V.~Tisserand,
A.~Zghiche
\inst{Laboratoire de Physique des Particules, F-74941 Annecy-le-Vieux, France }
A.~Palano,
A.~Pompili
\inst{Universit\`a di Bari, Dipartimento di Fisica and INFN, I-70126 Bari, Italy }
% ...

\end{center}\newpage

%% file: pubboard/acknowledgements.tex
We are grateful for the 
extraordinary contributions of our \pep2\ colleagues in
achieving the excellent luminosity and machine conditions
that have made this work possible.
The success of this project also relies critically on the 
expertise and dedication of the computing organizations that 
support \babar.
The collaborating institutions wish to thank 
SLAC for its support and the kind hospitality extended to them. 
This work is supported by the
US Department of Energy
and National Science Foundation, the
Natural Sciences and Engineering Research Council (Canada),
Institute of High Energy Physics (China), the
Commissariat \`a l'Energie Atomique and
Institut National de Physique Nucl\'eaire et de Physique des Particules
(France), the
Bundesministerium f\"ur Bildung und Forschung and
Deutsche Forschungsgemeinschaft
(Germany), the
Istituto Nazionale di Fisica Nucleare (Italy),
the Foundation for Fundamental Research on Matter (The Netherlands),
the Research Council of Norway, the
Ministry of Science and Technology of the Russian Federation, and the
Particle Physics and Astronomy Research Council (United Kingdom). 
Individuals have received support from 
the A. P. Sloan Foundation, 
the Research Corporation,
and the Alexander von Humboldt Foundation.